\documentclass[prd, floatfix, twocolumn, preprintnumbers, letterpaper, superscriptaddress, nofootinbib]{revtex4-1}

\usepackage{amsfonts, amsmath, amssymb, latexsym, graphicx, tikz, float, subfigure, epstopdf}

\usepackage[colorlinks=true, citecolor=blue, linkcolor=blue, urlcolor=black]{hyperref}

\usepackage{dcolumn, psfrag, wrapfig, makeidx, color, multirow}

\newcommand{\be}{\begin{equation}}
\newcommand{\ee}{\end{equation}}

\graphicspath{{Images/}}

\definecolor{lime}{HTML}{A6CE39}
\newcommand{\orcidicon}{%
    \begin{tikzpicture}
    \draw[lime, fill=lime] (0,0) circle [radius=0.16] node[white] {{\fontfamily{qag}\selectfont \tiny ID}};
    \draw[white, fill=white] (-0.0625,0.095) circle [radius=0.007];
    \end{tikzpicture} \hspace{-2mm}
}
\newcommand\orcidMahdi{{\href{https://orcid.org/0000-0003-1196-9493}{\orcidicon}}}

\begin{document}

\title{Quasinormal modes as probes of black hole reentrant phase transitions}

\author{M. Kord Zangeneh\orcidMahdi}
\email{mkzangeneh@scu.ac.ir}
\affiliation{Physics Department, Faculty of Science, Shahid Chamran University of Ahvaz, Ahvaz 61357-43135, Iran}

\author{S. Abdollahi}
\affiliation{Physics Department, Faculty of Science, Shahid Chamran University of Ahvaz, Ahvaz 61357-43135, Iran}

\author{D. Afshar}
\affiliation{Physics Department, Faculty of Science, Shahid Chamran University of Ahvaz, Ahvaz 61357-43135, Iran}

\begin{abstract}
We investigate the interplay between thermodynamic phase transitions and the dynamical behavior of quasinormal modes (QNMs) in weakly nonlinear charged anti-de Sitter black holes. In a non-extended phase space, we identify a reentrant large-small-large(intermediate) phase transition for these black holes. Employing the shooting method to solve the Klein-Gordon equation for a massless scalar field, we compute QNM frequencies across these phases, revealing distinct behaviors in different phases. Specifically, to the best of our knowledge, this is the first report of utilizing QNMs as dynamical probes of thermodynamic phases during reentrant phase transitions including zeroth and first-order transitions. These findings establish QNMs as a powerful dynamical probe for thermodynamic phase transitions, providing new insights into the thermodynamic and dynamical properties of nonlinear charged black holes as well.
\end{abstract}

\maketitle

\section{Introduction}

In the early 1970s, approximately sixty years after the introduction of the black hole solution, physicists established a connection between black holes and thermodynamics, leading to the formulation of the laws of black hole thermodynamics \cite{Hawking:1971tu, Bekenstein:1973ur, Bardeen:1973gs, Hawking:1975vcx}. This remarkable discovery highlighted the profound relationship between the principles of black hole physics and thermodynamic laws, marking a significant advancement in theoretical physics.

Following this, researchers introduced phase transitions analogous to thermodynamic phase transitions in the context of black holes. Hawking and Page demonstrated the existence of a first-order phase transition between thermal radiation and a black hole in anti-de Sitter (AdS) spacetime \cite{Hawking:1982dh}. Witten later linked this phase transition to one occurring in quark-gluon plasma \cite{Witten:1998zw}. Further studies revealed that, similar to the Van der Waals gas-liquid phase transition, the charged AdS black hole undergoes isobaric and isothermal first-order phase transitions between large and small black holes within the extended thermodynamic phase space \cite{Chamblin:1999tk, Dolan:2010ha, Kubiznak:2012wp, Gunasekaran:2012dq}. In this extended phase space, the variable cosmological constant is treated as pressure, while its conjugate quantity is the system's volume.

In the extended phase space approach, it is crucial to connect the pressure of the black hole to the cosmological constant, a quantity associated with space, thus extending the phase space. Alternatively, the cosmological constant can be treated as a fixed value while allowing charge to vary \cite{Dehyadegari:2016nkd}. From this perspective, the cosmological constant is not viewed as a variable thermodynamic quantity; instead thermodynamic processes occur within a phase space where the black hole evolves in a fixed AdS background geometry. This model can effectively serve as a substitute for the extended phase space. Moreover, treating the black hole charge as a variable is a natural choice, and it provides a more accurate description of phase transitions and critical behavior \cite{Dehyadegari:2016nkd}. This allows the corresponding response function to effectively characterize the system's stability and instability \cite{Dehyadegari:2016nkd}. This approach has been applied to study the phase transition of the nonlinearly charged AdS black hole, revealing the occurrence of a reentrant phase transition (RPT) \cite{Dehyadegari:2017hvd}.

It has been shown that nonlinear electrodynamic models effectively address several limitations inherent in Maxwell linear electrodynamics. While Maxwell theory has proven remarkably successful, it encounters significant issues when analyzing point-charged particles, leading to unacceptable results. To overcome these challenges, Born and Infeld introduced the first nonlinear electrodynamics model in 1934 to resolve such discrepancies \cite{born1934quantum,born1934foundations}. Notably, these nonlinear theories have effectively tackled problems such as the infinite self-energy of point charges and the severe field divergences near point charges. They also provide solutions to a well-known issue in gravitational theory: the singularities associated with the big bang and black holes. Studies indicate that nonlinear electrodynamic theories can eliminate these singularities \cite{ayon1999non,ayon1999new,de2002nonlinear,irina2004regular,
corda2010removing,corda2011inflation}.

In the realm of nonlinear electrodynamics, most Lagrangians exhibit a similar structure in the weakly nonlinear regime, often referred to as weakly nonlinear electrodynamics. The Lagrangians of various nonlinear electrodynamic theories, such as Born-Infeld \cite{born1934quantum,born1934foundations}, logarithmic \cite{soleng1995charged}, power-law \cite{hassaine2007higher}, and exponential \cite{hendi2012asymptotic} electrodynamics, each based on different motivations, converge to the following form in the weakly nonlinear approximation \cite{hendi2015thermodynamic}: 
\begin{equation}
L (F) = -F + \alpha F^{2} + O(\alpha^{2}). \label{eq1}
\end{equation}
Here, $\alpha$ represents the nonlinear parameter that controls deviations from Maxwell theory, $F = F_{\mu \nu} F^{\mu \nu}$ is the Maxwell invariant, and $F_{\mu \nu} = \partial_{\mu} A_{\nu} - \partial_{\nu} A_{\mu}$ denotes the electromagnetic field tensor, with $A_{\mu}$ as the gauge potential.

Reentrant phase transitions refer to situations in which two or more distinct phase transitions occur as a thermodynamic control parameter changes monotonically, such that the system ultimately returns to a macroscopic phase identical to its initial state. This phenomenon has been observed in various physical systems, including nicotine/water mixtures \cite{Nwater}, granular superconductors, liquid crystals, binary gases, ferroelectrics, and gels (see \cite{RPTreview} and references therein). In recent years, RPTs have attracted considerable attention in the context of black hole thermodynamics \cite{KerrRPT, MRPT, hairyRPT, dSRPT, Zou:2016sab, MicroscopicRPT,Dehghani:2020blz,Liu:2022spy,Bai:2022vmx,Frassino:2023wpc}. In such gravitational systems, the reentrant behavior typically manifests as a sequence of large-small-large(intermediate) black hole phases. Remarkably, this process is accompanied by a finite jump in the Gibbs free energy, signaling the presence of a zeroth-order phase transition.

Some systems exhibit individual modes. In non-dissipative systems, these modes are real and referred to as normal modes. In contrast, black holes are considered dissipative systems due to their radiation, which leads to the emergence of imaginary modes when the black hole is perturbed. These imaginary modes indicate energy loss and are called quasinormal modes (QNMs). The study of black hole QNMs began in the early 1970s \cite{vishveshwara1970scattering, press1971long, Chandrasekhar:1975zza}. Since then, various aspects of black hole QNMs have been extensively investigated, establishing QNMs as powerful probes of black hole properties \cite{Nollert:1999ji, Kokkotas:1999bd, Wang:2005vs, Fernando:2005bc, Konoplya:2011qq,Konoplya:2002zu,Konoplya:2002ky,Cardoso:2003cj,Mahapatra:2016dae,Lin:2024ubg,Ribeiro:2025amp}. Furthermore, the effects of the small-large black hole phase transition on QNMs have been studied for charged AdS black holes within the framework of the extended phase space \cite{Liu:2014gvf}. Recently, the interplay between QNM behaviors and thermodynamic phase transitions has been studied, particularly in relation to hairy black holes in Einstein–Maxwell–scalar gravity \cite{Priyadarshinee:2023exb} and regular hairy black holes of the Bardin and Hayward types \cite{Guo:2024jhg}. Additionally, the QNMs of charged black holes in F(R)–Euler–Heisenberg gravity have been shown to encode information about thermodynamic phase transitions, revealing a correspondence between variations in the QNM spectrum and changes in the thermodynamic phase structure \cite{Hou:2025bli}.

In this paper, we investigate the relationship between the isocharge phase transition in a non-extended phase space for weakly nonlinear charged AdS black holes and the behavior of their QNMs. We demonstrate that, even within this alternative perspective on the thermodynamic phase space of black holes, the behavior of QNMs can still reflect the thermodynamic phases and transitions of the black holes. This finding reaffirms the strong connection between the dynamical and thermodynamic behaviors of black holes. In particular, as far as we are aware, this is the first report indicating that QNMs could serve as dynamical probes of thermodynamic phases during reentrant phase transitions, including zeroth-order phase transitions.

The structure of this paper is as follows: In Section \ref{sec2}, we review the thermodynamics of weakly nonlinear charged AdS black holes. Section \ref{phtr} explores the isocharge phase transition of these black holes in a non-extended phase space. In Section \ref{sec4}, we numerically calculate the QNMs of a scalar field perturbing the black holes and examine the relationship between the behavior of the QNM frequencies and the thermodynamic phase transitions. Finally, the last section summarizes our findings and presents concluding remarks.

\section{Thermodynamics of Weakly Nonlinear Charged AdS Black Holes \label{sec2}}

\begin{figure*}[t]
    \subfigure[~\( G-T \)]{\includegraphics[width=0.48\textwidth]{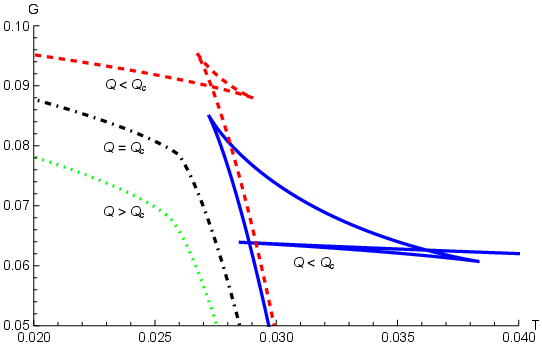}}\quad
    \subfigure[~\( T-r_H \)]{\includegraphics[width=0.48\textwidth]{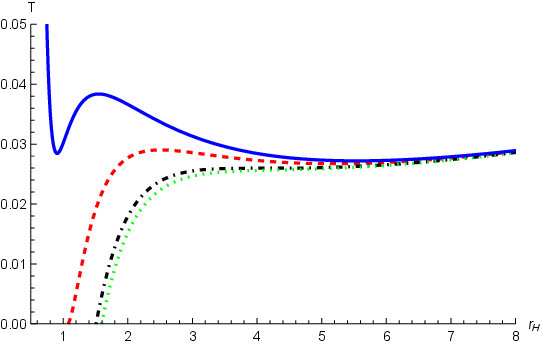}}
    \caption{(a) Gibbs free energy as a function of temperature. The black dot-dashed curve represents the critical charge \(Q_c = 0.133\) case. The red dashed and blue solid curves depict phase transitions at charges below the critical value, specifically \(Q = 0.105\) and \(Q = 0.0715\), respectively. The green dotted curve indicates the absence of phase transition for charges above the critical charge (\(Q > Q_c\)). (b) Corresponding \( T-r_H \) diagram.}
    \label{fig1}
\end{figure*}

We consider a four-dimensional Einstein gravity theory coupled to weakly nonlinear electrodynamics with a negative cosmological constant \(\Lambda = -3/l^2\), where \(l\) is the AdS radius. The action is given by \cite{hendi2015thermodynamic}:
\begin{equation}
I = -\frac{1}{16\pi} \int d^4 x \sqrt{-g} \left[ R - 2\Lambda + L(F) \right],
\end{equation}
where \(R\) is the Ricci scalar, \(g\) is the metric determinant, and
\( L(F) \) is the Lagrangian of weakly nonlinear electrodynamics given by Eq. (\ref{eq1}). Varying the action with respect to the metric \(g_{\mu\nu}\) and the gauge potential \(A_{\mu}\) yields the field equations:
\begin{align}
&G_{\mu\nu} + \Lambda g_{\mu\nu} = \frac{1}{2} g_{\mu\nu} L(F) - 2 L_F F_{\mu\lambda} F_{\nu}^{\lambda}, \label{eq:field_eq1} \\
&\partial_{\mu} (\sqrt{-g} L_F F^{\mu\nu}) = 0, \label{eq:field_eq2}
\end{align}
in which \(G_{\mu\nu}\) is the Einstein tensor and \(L_F = \partial L / \partial F \). The solution to these equations is a weakly nonlinear charged AdS black hole with the metric \cite{hendi2015thermodynamic}:
\begin{align}
ds^2 &= -f(r) dt^2 + \frac{dr^2}{f(r)} + r^2 \left(d\theta^2 + \sin^2\theta d\phi^2\right), \label{eq:metric} \\
f(r) &= 1 - \frac{m}{r} + \frac{r^2}{l^2} + \frac{q^2}{r^2} - \frac{2 q^4 \alpha}{5 r^6} + O\left(\alpha^2 \right), \label{eq:metric_f} \\
F_{tr} &= \frac{q}{r^2} - 4 \left(\frac{q}{r^2}\right)^3 \alpha + O\left(\left(\frac{q}{r^2}\right)^5 \alpha^2\right), \label{eq:field_strength}
\end{align}
where \(m\) and \(q\) are integration constants related to the black hole mass \(M = m / 8\pi\) and electric charge \(Q = q / 4\pi\), {with \(M\) and \(Q\) defined per unit volume. The corresponding volume in our setup is that of the unit \(2\)-sphere, \(V_2=4\pi\).}

Setting \(f(r_H) = 0\) at the event horizon radius \(r_H\), the mass is:
\begin{equation}
M = \frac{r_H}{8 \pi} + \frac{r_H^3}{8 \pi l^2} + \frac{2\pi Q^2}{r_H} - \frac{64\pi^3 Q^4 \alpha}{5 r_H^5} + O(\alpha^2). \label{eq:mass}
\end{equation}
The Hawking temperature, derived from the surface gravity, is:
\begin{equation}
T = \frac{f'(r_H)}{4\pi} = \frac{1}{4 \pi r_H} + \frac{3 r_H}{4 \pi l^2} - \frac{4 \pi Q^2}{r_H^3} + \frac{128 \pi^3 Q^4 \alpha}{r_H^7} + O(\alpha^2), \label{eq:temperature}
\end{equation}
and the entropy {per unit volume}, proportional to the horizon area, is:
\begin{equation}
S = \frac{r_H^2}{4}. \label{eq:entropy}
\end{equation}
The Gibbs free energy is a central concept in the analysis of phase transitions, as it governs both their occurrence and their classification by order. Discontinuities in the Gibbs free energy itself as a function of temperature indicate a zeroth-order phase transition, whereas discontinuities in its first derivative with respect to temperature correspond to a first-order transition. Similarly, higher-order phase transitions are characterized by discontinuities or divergences in higher derivatives of the Gibbs free energy. The Gibbs free energy {per unit volume}, defined as \(G = M - T S\), is:
\begin{equation}
G = \frac{r_H}{16\pi} - \frac{r_H^3}{16 \pi l^2} + \frac{3 \pi Q^2}{r_H} - \frac{224\pi^3 Q^4 \alpha}{5 r_H^5} + O(\alpha^2). \label{eq:gibbs}
\end{equation}
These quantities govern the thermodynamic behavior of weakly nonlinear charged AdS black holes, with the nonlinear parameter \(\alpha\) introducing corrections that influence phase transitions and stability, as explored in the next section.

\begin{figure*}[t]
    \subfigure[~\( G-T \)]{\includegraphics[width=0.31\textwidth]{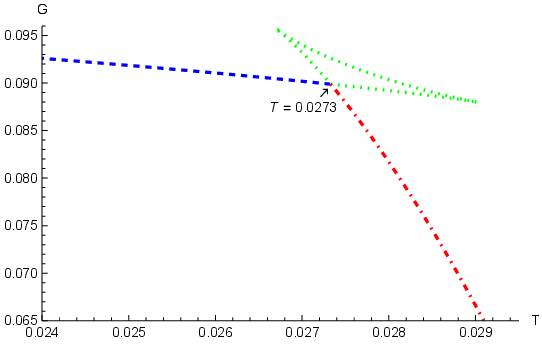} \label{fig2a}}\quad
    \subfigure[~\( T-r_H \)]{\includegraphics[width=0.31\textwidth]{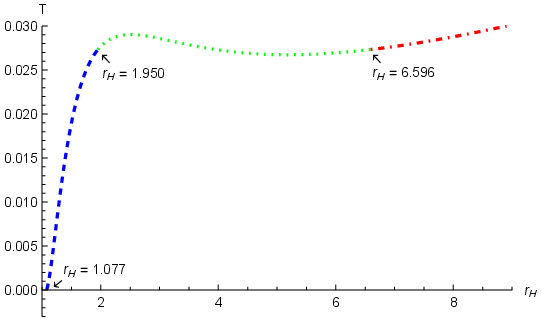} \label{fig2b}}
        \subfigure[~\( M-r_H \)]{\includegraphics[width=0.31\textwidth]{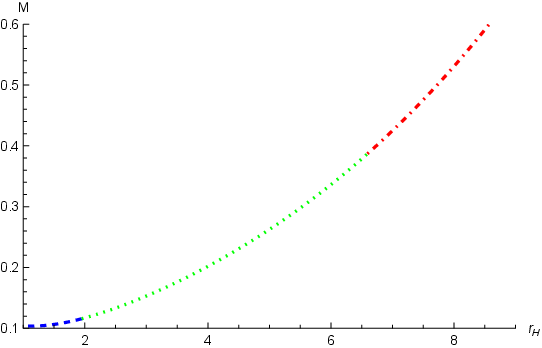} \label{fig2c}}
    \caption{(a) Gibbs free energy as a function of temperature for \(Q = 0.105 < Q_c\) which exhibits a small-large black hole phase transition at \(T = 0.0273\). (b) Corresponding \( T-r_H \) diagram (c) Corresponding \( M-r_H \) diagram.}
    \label{fig2}
\end{figure*}

\begin{figure*}[t]
    \subfigure[~\( G-T \)]{\includegraphics[width=0.31\textwidth]{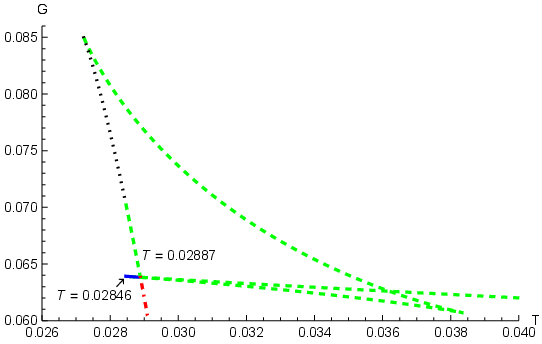} \label{fig3a}}\quad
    \subfigure[~\( T-r_H \)]{\includegraphics[width=0.31\textwidth]{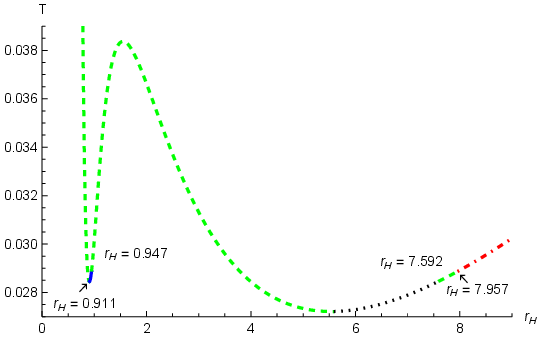} \label{fig3b}}
        \subfigure[~\( M-r_H \)]{\includegraphics[width=0.31\textwidth]{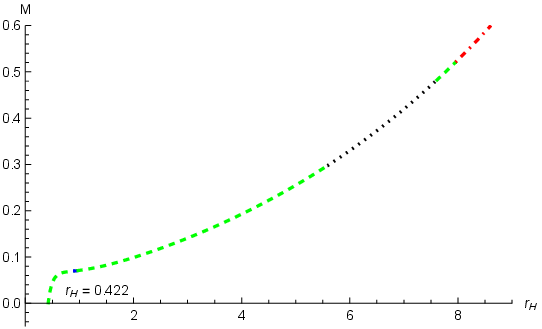} \label{fig3c}}
    \caption{(a) Gibbs free energy as a function of temperature for \(Q = 0.0715 < Q_c\) which exhibits a zeroth-order phase transition at \(T = 0.02846\) and a first-order phase transition at \(T = 0.02887\). (b) Corresponding \( T-r_H \) diagram. (c) Corresponding \( M-r_H \) diagram.}
    \label{fig3}
\end{figure*}

\section{Phase transitions in a non-extended phase space \label{phtr}}

As discussed earlier in the introduction, the small-large black hole phase transition was initially investigated within the extended phase space framework, where the cosmological constant is interpreted as a thermodynamic pressure \cite{Chamblin:1999tk, Dolan:2010ha, Kubiznak:2012wp, Gunasekaran:2012dq}. In contrast, an alternative approach considers the cosmological constant as a fixed parameter, allowing the black hole charge to vary in a non-extended phase space \cite{Dehyadegari:2016nkd}. This perspective is particularly advantageous for analyzing isocharge processes, in which the charge remains constant while the event horizon radius varies and different charge values determine the occurrence and type/order of phase transition. Such a framework facilitates a detailed exploration of critical phenomena and thermodynamic stability \cite{Dehyadegari:2017hvd}.

In the isocharge process, the critical point, marking the onset of phase transitions, is determined by the conditions \cite{Dehyadegari:2016nkd}:
\begin{equation}
\frac{\partial T}{\partial r_H} = 0 = \frac{\partial^2 T}{\partial r_H^2}. \label{eq:critical_conditions}
\end{equation}
Solving the critical conditions, we obtain the critical parameters to first order of nonlinear parameter $\alpha$:
\begin{align}
r_{Hc} &= \frac{l}{\sqrt{6}} - \frac{7}{3 \sqrt{6} l} \alpha + O(\alpha^2), \label{eq:rHc} \\
Q_c &= \frac{l}{24 \pi} + \frac{7}{72 \pi l} \alpha + O(\alpha^2), \label{eq:Qc} \\
T_c &= \frac{1}{\pi l} \sqrt{\frac{2}{3}} - \frac{1}{3 \pi l^3} \sqrt{\frac{2}{3}} \alpha + O(\alpha^2). \label{eq:Tc}
\end{align}
These parameters recover the Reissner-Nordström-AdS critical point at \(\alpha = 0\) \cite{Kubiznak:2012wp} and quantify the nonlinear corrections.

The heat capacity {per unit volume} at constant charge, \(C_Q = T (\partial S / \partial T)_Q\), determines local thermodynamic stability:
\begin{equation}
C_Q = T \frac{\partial S}{\partial r_H} \left( \frac{\partial T}{\partial r_H} \right)^{-1} = T \frac{r_H}{2} \left( \frac{\partial T}{\partial r_H} \right)^{-1}, \label{eq:heat_capacity}
\end{equation}
where positive (negative) \(C_Q\) indicates stable (unstable) phases, corresponding to positive (negative) slopes in the \(T\)-\(r_H\) diagram.

In Figure \ref{fig1}, the behavior of the Gibbs free energy \(G\) as a function of temperature \(T\) is illustrated for various charge values \(Q\). The corresponding relationship between temperature and horizon radius \(r_H\) is also shown. Notably, for \(Q < Q_c \approx 0.133\), the system exhibits two types of phase transitions: zeroth-order and first-order. In this figure and in the subsequent analysis, we fix the AdS radius at \(l = 10\) and the nonlinear parameter at \(\alpha = 0.12\), arbitrarily. The latter choice ensures the validity of the weakly nonlinear approximation as it keeps the $n(\geq 2)$-th order nonlinear correction terms subdominant compared to the Maxwell term. This requires \(\left(\alpha q^2 / r_H^4 \right)^n \ll 1\) (see Eq. (\ref{eq:field_strength})) \cite{hendi2015thermodynamic}. For the parameter ranges considered in this work, this condition is satisfied.

In Figure \ref{fig2}, the plots of \(G\) versus \(T\), \(T\) versus \(r_H\) {and \(M\) versus \(r_H\)} are shown for \(Q = 0.105\), which is below the critical charge \(Q_c\). In this regime, the system undergoes a first-order phase transition, characterized by minimization of the Gibbs free energy through a transition from a small to a large black hole phase at \(T = 0.0273\). As depicted in Fig. \ref{fig2b}, the positive slope of the \(T - r_H\) curve in both phases indicates that the system is thermodynamically stable in small and large phases. {We restrict the analysis to \(r_H > 1.077\), for which both the mass and temperature remain positive. In this range, \(r_H\) represents the outermost (largest) root of the metric function, although additional smaller roots may exist. This is consistent with the \(M-r_H\) relation shown in Fig. \ref{fig2c}, where \(M(r_H)\) is an increasing and one-to-one function over the considered range. At the lower bound, \(r_H=1.077\), \(M(r_H)\) attains a local minimum, since \(T=\partial M/\partial S \propto \partial M/\partial r_H\) vanishes there and remains positive for larger values of \(r_H\), as exhibited in Fig. \ref{fig2b}.}

Figure \ref{fig3} presents a particularly intriguing case at \(Q = 0.0715 < Q_c\), where the system experiences reentrant large-small-large(intermediate) phase transition. The large-small phase transition is a first-order phase transition. The zeroth-order phase transition, evidenced by a discontinuity in the Gibbs free energy at \(T = 0.02846\), is a relatively rare phenomenon in black hole thermodynamics. It involves an abrupt jump between small and large(intermediate) black hole phases, as previously reported in certain nonlinear electrodynamics models \cite{Dehyadegari:2017hvd}. As seen in Fig. \ref{fig3b}, the positive slopes of the \(T - r_H\) curves in small and large phases further confirm the thermodynamic stability of the system in these regions. {In this case, the temperature remains positive throughout the considered range, as exhibited in Fig. \ref{fig3b}, while restricting the analysis to \(r_H > 0.422\) ensures that the mass remains positive, as illustrated in Fig. \ref{fig3c}. Moreover, \(M(r_H)\) is a monotonically increasing and one-to-one function over this range, since \(T \propto \partial M/\partial r_H\) remains positive. Therefore, for each allowed value of the mass, there is a unique corresponding value of \(r_H\), implying that the metric function has only one root, which corresponds to the event horizon.}

These phase transitions highlight the rich thermodynamic structure of weakly nonlinear charged AdS black holes, with nonlinear electrodynamics introducing novel critical phenomena not present in the linear Maxwell theory \cite{Rosh:2025}.

Subsequently, we will investigate the behavior of QNMs during phase transitions and demonstrate how their signatures manifest in both zeroth-order and first-order transitions.

\begin{figure*}[t]
    \subfigure[~Small black hole phase]{\includegraphics[width=0.48\textwidth]{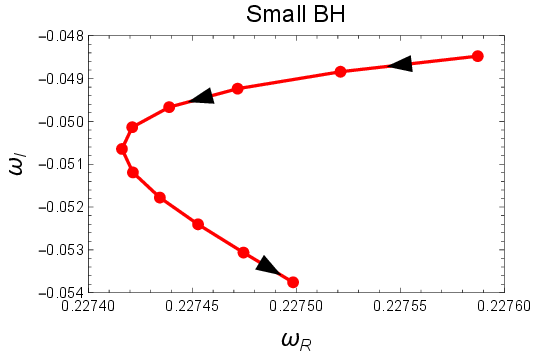} \label{fig4a}}\quad
    \subfigure[~Large black hole phase]{\includegraphics[width=0.48\textwidth]{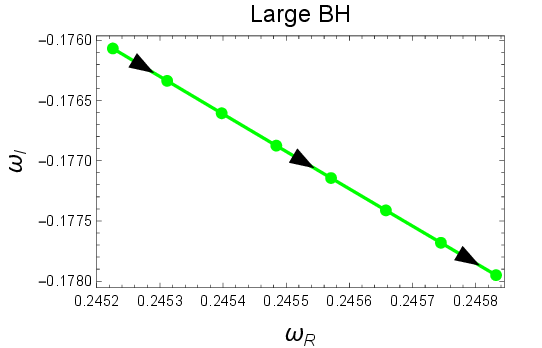} \label{fig4b}}
\caption{Imaginary part of QNM frequencies versus real part for the first-order phase transition. The arrows indicate the increase of black hole size.}
\label{fig4}
\end{figure*}

\section{QNMs and phase transitions \label{sec4}}

\begin{table}[t]
\centering
\begin{tabular}{cccc}
\hline
\textbf{$r_H$} & \textbf{$T$} & \textbf{$\omega_R$} & \textbf{$\omega_I$} \\
\hline
1.590 & 0.0216 & 0.227521 & -0.048841 \\
1.650 & 0.0231 & 0.227453 & -0.049447 \\
1.710 & 0.0243 & 0.227421 & -0.050136 \\
{1.750} & {0.0250} & {0.227416} & {-0.050644} \\
1.770 & 0.0253 & 0.227418 & -0.050913 \\
1.830 & 0.0261 & 0.227434 & -0.051779 \\
1.890 & 0.0268 & 0.227463 & -0.052729 \\
1.950 & 0.0273 & 0.227498 & -0.053757 \\
\hline
6.596 & 0.0273 & 0.245225 & -0.176068 \\
6.716 & 0.0274 & 0.246272 & -0.179294 \\
6.836 & 0.0275 & 0.247348 & -0.182518 \\
6.956 & 0.0276 & 0.248454 & -0.185741 \\
7.076 & 0.0277 & 0.249591 & -0.188964 \\
\hline
\end{tabular}
\caption{QNM frequencies for small (above the line) and large (below the line) black hole phases during a first-order phase transition with \(Q=0.105\).}
\label{table1}
\end{table}

To explore the dynamical signatures of phase transitions discussed in section \ref{phtr}, we analyze the QNMs associated with a massless scalar field perturbing the black hole spacetime. The evolution of the scalar field is governed by the Klein-Gordon wave equation in a curved background:
\be
\partial_{\mu}\left(\sqrt{-g} g^{\mu\nu} \partial_{\nu}\right) \psi = 0.
\ee
Applying separation of variables, the wave function can be written as \(\psi(t, r, \theta, \phi) = e^{-i \omega t} R(r) Y(\theta, \phi)\), where \(Y(\theta, \phi)\) denote spherical harmonics and \(\omega\) is the complex QNM frequency, yields a radial differential equation:
\be
R''(r) + \left[\frac{f'(r)}{f(r)} + \frac{2}{r}\right] R'(r) + \frac{\omega^2 R(r)}{f(r)^2} = 0,
\ee
where \(f(r)\) is specified by Eq. \eqref{eq:metric_f}. Near the event horizon, where \(f(r_H) = 0\), the radial equation simplifies to:
\be
R''(r) + \frac{f'(r)}{f(r)} R'(r) + \frac{\omega^2}{f(r)^2} R(r) = 0,
\ee
which admits the solution \( R(r) = \exp \left(-i \int \left[\omega/f(r)\right] dr \right) \) representing purely ingoing waves at the horizon, consistent with the causal boundary condition that no outgoing waves escape from inside the horizon. To facilitate numerical integration, we redefine the radial function as:
\be
R(r) = \tilde{R}(r) \exp \left(-i \int \frac{\omega}{f(r)} dr \right),
\ee
leading to a transformed radial equation:
\be
\tilde{R}''(r) + \left[\frac{f'(r)}{f(r)} - \frac{2 i \omega}{f(r)} + \frac{2}{r}\right] \tilde{R}'(r) - \frac{2 i \omega^2}{r f(r)} \tilde{R}(r) = 0.
\ee
We solve this differential equation numerically via the shooting method, imposing boundary conditions that \(\tilde{R}(r_H) = 1\) at the horizon, and \(\tilde{R}(r) \to 0\) as \(r \to \infty\), which captures that the wave function is localized and the expected decay of perturbations. The complex QNM frequencies \(\omega = \omega_R + i \omega_I\) encode the oscillation and damping characteristics of the perturbations: \(\omega_R\) indicates the oscillation frequency, and \(\omega_I\) corresponds to the damping rate, with \(\omega_I < 0\) signifying stability through exponential decay of perturbations.

The iterative adjustment of \(\omega\) in the shooting method ensures that the boundary conditions are satisfied. This numerical approach has been employed in analyzing QNMs in various black hole backgrounds, providing critical insights into their stability and dynamic response \cite{Liu:2014gvf, Priyadarshinee:2023exb, Rosh:2025}.

In this section, we numerically compute the QNM frequencies associated with the identified phase transitions. The results are presented for two specific cases, corresponding to the phase transitions discussed in section \ref{phtr}.


\begin{figure*}[t]
    \subfigure[~Large black hole phase]{\includegraphics[width=0.31\textwidth]{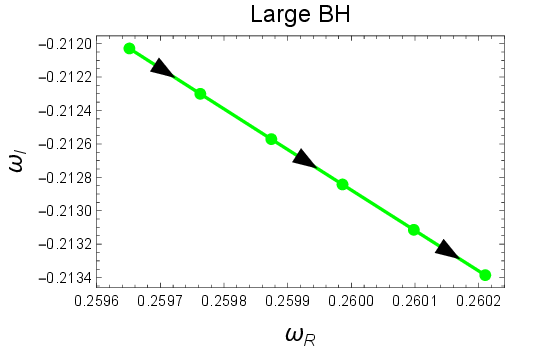} \label{fig5a}}
    \subfigure[~Small black hole phase]{\includegraphics[width=0.31\textwidth]{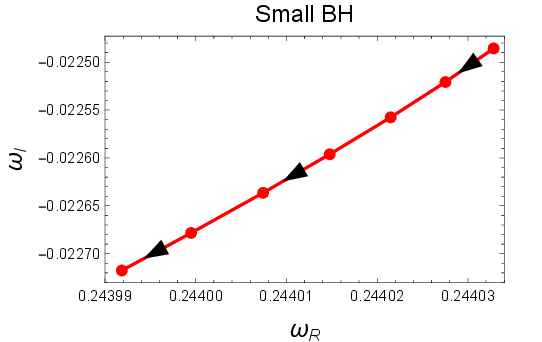} \label{fig5b}}\quad
    \subfigure[~Large(intermediate) black hole phase]{\includegraphics[width=0.31\textwidth]{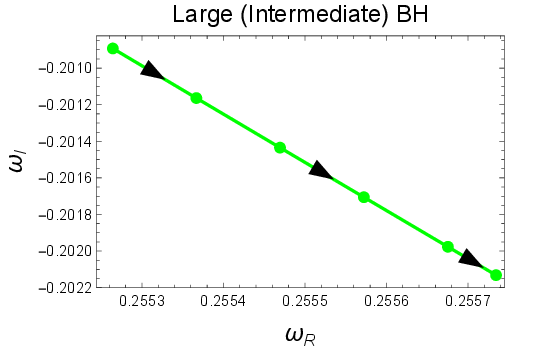} \label{fig5c}}
\caption{Imaginary part of QNM frequencies versus real part for the reentrant large-small-large(intermediate) phase transition. The arrows indicate the increase of black hole size.}
\label{fig5}
\end{figure*}

\begin{table}[t]
\centering
\begin{tabular}{cccc}
\hline
\textbf{$r_H$} & \textbf{$T$} & \textbf{$\omega_R$} & \textbf{$\omega_I$} \\
\hline
7.997 & 0.02892 & 0.260098 & -0.213114 \\
7.987 & 0.02891 & 0.259986 & -0.212843 \\
7.977 & 0.02889 & 0.259874 & -0.212571 \\
7.967 & 0.02888 & 0.259763 & -0.212300 \\
7.957 & 0.02887 & 0.259651 & -0.212029 \\
\hline
0.947 & 0.02887 & 0.243992 & -0.022718 \\
0.941 & 0.02877 & 0.243999 & -0.022678 \\
0.929 & 0.02860 & 0.244015 & -0.022596 \\
0.917 & 0.02849 & 0.244028 & -0.022521 \\
0.911 & 0.02846 & 0.244033 & -0.022486 \\
\hline
7.592 & 0.02846 & 0.255734 & -0.202131 \\
7.587 & 0.02845 & 0.255675 & -0.201976 \\
7.577 & 0.02844 & 0.255572 & -0.201706 \\
7.567 & 0.02843 & 0.255469 & -0.201435 \\
7.557 & 0.02842 & 0.255367 & -0.201164 \\
\hline
\end{tabular}
\caption{QNM frequencies for large, small and intermediate(large) black hole phases (from above to down) during a reentrant phase transition with \(Q=0.0715\).}
\label{table2}
\end{table}

\subsection{First-order phase transition \label{foptsubsec}}


For \(Q=0.105 < Q_c \approx 0.133\), a first-order phase transition occurs at  \(T = 0.0273\) (see Fig. \ref{fig2a}). Table \ref{table1} lists the computed QNM frequencies for both the small and large black hole phases across this transition. In the small black hole phase, the real part \(\omega_R\) initially decreases with increasing horizon radius \(r_H\), indicating a reduction in oscillation frequency and mode energy, then increases at larger \(r_H\). {This turn-over occurs at \(r_H=1.750\), where \(\omega_R\) reaches its minimum value.} The magnitude of the imaginary part \(|\omega_I|\) systematically increases, implying that perturbations decay more rapidly as the black hole grows. Conversely, in the large black hole phase, both \(\omega_R\) and \(|\omega_I|\) increase with \(r_H\), reflecting higher vibrational energies and faster damping rates for larger black holes.

These behaviors are visually summarized in the \(\omega_I - \omega_R\) plots of Fig. \ref{fig4}, where the red curve (small black holes) and green curve (large black holes) display distinct trends, highlighting the sensitivity of QNMs to the underlying thermodynamic phase. {In particular, within the small black hole phase, for smaller values of the horizon radius, the \(\omega_I-\omega_R\) trajectory exhibits a behavior distinct from that of the large black hole phase, indicating a correlation between the QNM behavior and the corresponding thermodynamic phase.} The AdS black hole in the presence of linear Maxwell electrodynamics shows the same behavior \cite{Rosh:2025}, so the present result reconfirms that the QNMs {can provide a signature of the thermodynamic phases of black holes.}

\subsection{Reentrant phase transition \label{rptsubsec}}

For \(Q=0.0715\), we observe a reentrant phase transition, including a large-small first-order phase transition and a small-large(intermediate) zeroth-order phase transition, occurring at \(T = 0.02887\) and \(T = 0.02846\) (see Fig. \ref{fig3a}). The corresponding QNM frequencies are detailed in Table \ref{table2}. In the large black hole phase, both \(\omega_R\) and \(|\omega_I|\) decrease with \(r_H\), reflecting lower-energy oscillations that decay more slowly as the black hole size shrinks. Conversely, in the small black hole phase, \(\omega_R\) increases as \(r_H\) decreases, indicating an increase in oscillation frequencies and mode energies. Concurrently, \(|\omega_I|\) decreases, signifying a slower damping rate. The behavior of QNM frequencies in the intermediate black hole phase is the same as in the large black hole phase. This shows that during zeroth- and first-order phase transitions, we can infer the phase and the phase transition by studying the behavior of QNM frequencies.

Fig. \ref{fig5} shows the trajectories in the \(\omega_I-\omega_R\) plane for these transitions. The red curve corresponds to small black holes, while the green curves correspond to large and intermediate black holes, highlighting the markedly different phase-dependent behavior of QNMs. Notably, the slopes of these trajectories have opposite signs during each transition, illustrating how thermodynamic phase influences dynamical properties. To the best of our knowledge, this is the first report suggesting that QNMs can serve as dynamical probes of thermodynamic phases during a reentrant phase transition, including zeroth- and first-order transitions. The consistently negative \(\omega_I\) across all cases studied in the present section confirms the dynamical stability of the system, since perturbations decay exponentially.

The distinct QNM behaviors align with the thermodynamic stability analysis, as stable phases characterized by positive specific heat (discussed in Sec. \ref{phtr}) correspond to regions of consistent frequency trends, while transitions mark abrupt changes in dynamical properties. This correlation suggests that QNMs can serve as probes of thermodynamic phase transitions in non-extended phase space, similar to their role in extended phase space studies \cite{Liu:2014gvf, Priyadarshinee:2023exb, Guo:2024jhg}. Notably, the signature of the zeroth-order transition in the QNM spectrum underscores the sensitivity of dynamical responses to rare thermodynamic phenomena, opening new avenues for exploring the physics of black holes.

\section{Summary and Concluding remarks}

This study analyzed the interplay between phase transitions in a non-extended phase space and QNM frequencies for weakly nonlinear charged AdS black holes. The critical parameters, derived to first order in the nonlinear parameter \(\alpha\), quantified the deviation from linear electrodynamics and highlighted the modified critical behavior. For charges below the critical value, we identified a first-order phase transition between large and small black holes as well as a reentrant phase transition that includes a first-order large-small transition and a zeroth-order small-large(intermediate) transition. This reveals a rich thermodynamic structure induced by nonlinear electrodynamics.

The numerical solutions of the Klein–Gordon equation, obtained using the shooting method for massless scalar field perturbations in black hole spacetime, yielded QNM frequencies that distinctly reflect the thermodynamic phases. During the first-order phase transition, in the small black hole phase, the real part of the QNM frequencies, \(\omega_R\), initially decreases with increasing horizon radius and then increases at larger black hole sizes, while the magnitude of the imaginary part \(|\omega_I|\) systematically increases. In the large black hole phase, both \(\omega_R\) and \(|\omega_I|\) increase with increasing black hole size. During the reentrant phase transition, in the large and intermediate black hole phases, both the real part and the magnitude of the imaginary part of the QNM frequencies increase with increasing black hole size. In the small black hole phase, \(\omega_R\) increases as the horizon radius decreases, whereas \(|\omega_I|\) decreases. The negative imaginary parts confirm dynamical stability, and the contrasting frequency trends in each phase underscore the potential of QNMs as a powerful tool for probing black hole thermodynamics.

These results extend earlier investigations of QNMs in extended phase space \cite{Liu:2014gvf, Priyadarshinee:2023exb, Guo:2024jhg} to the non-extended phase space, demonstrating the robustness of QNMs as reliable dynamical probes across different thermodynamic settings. As far as we are aware, this is the first report showing that QNMs can effectively diagnose thermodynamic phases in the presence of reentrant phase transitions, including both zeroth- and first-order transitions. This finding underscores the remarkable sensitivity of dynamical responses to rare thermodynamic phenomena and opens new avenues for exploring black hole physics. Moreover, the emergence of reentrant phase transitions emphasizes the distinctive influence of nonlinear electrodynamics in shaping black hole thermodynamics, with potential parallels in other nonlinear models \cite{Dehyadegari:2017hvd}.

{It is also worth emphasizing that the ability of QNMs to distinguish between different black hole phases is closely related to the underlying thermodynamic phase structure. In our previous study \cite{Rosh:2025}, which focused on linearly charged black holes exhibiting only a first-order phase transition, we found that at the critical charge, the distinction between the $\omega_I-\omega_R$ behavior of small and large black holes becomes apparent only sufficiently far from the phase transition point, within the small black hole phase, corresponding to smaller horizon radii. As the charge is decreased below the critical value, this distinction gradually becomes more pronounced, eventually extending throughout the entire small black hole phase. The results obtained in the present work are consistent with and further corroborate this behavior. On the other hand, for charge values above the critical charge, $Q>Q_c$, where no phase transition occurs, our investigations show that the slope of the $\omega_I-\omega_R$ curves remains negative. Consequently, the QNM spectrum does not provide a distinction between the small-like and large-like black hole configurations in this regime. These observations indicate that the QNM behavior is closely connected to the thermodynamic phase structure.}

The results may have broader implications for gauge/gravity duality \cite{Berti:2009kk}. The QNM frequencies could also inform gravitational wave signatures relevant for future observations \cite{Berti:2009kk, Jaramillo:2021tmt}. Furthermore, future research could explore QNMs corresponding to black holes in other nonlinear electrodynamic models or in higher dimensions, which is important from a string theory perspective, to further clarify the interplay between thermodynamics and dynamics. Investigating the effects of scalar field mass or spin on QNM signatures could provide deeper insights into the universality of these findings. Additionally, the study could be extended to other perturbations, such as vector and spinor modes. This work lays the groundwork for such explorations, reinforcing the profound connection between black hole dynamics and thermodynamics.

\acknowledgments{We thank the referee for constructive comments which helped us improve the paper significantly. MKZ thanks Shahid Chamran University of Ahvaz, Iran for supporting this work under research grant No. SCU.SP1404.37271. DA acknowledges Shahid Chamran University of Ahvaz, Iran for supporting this work under research grant No. SCU.SP1404.812.}


\begin{thebibliography}{99}

\bibitem{Hawking:1971tu}
S.~W.~Hawking,
Gravitational radiation from colliding black holes,
\hypersetup{urlcolor=blue}\href{https://doi.org/10.1103/PhysRevLett.26.1344}{Phys. Rev. Lett. \textbf{26}, 1344-1346 (1971)}.
\bibitem{Bekenstein:1973ur}
J.~D.~Bekenstein,
Black holes and entropy,
\hypersetup{urlcolor=blue}\href{https://doi.org/10.1103/PhysRevD.7.2333}{Phys. Rev. D \textbf{7}, 2333-2346 (1973)}.
\bibitem{Bardeen:1973gs}
J.~M.~Bardeen, B.~Carter and S.~W.~Hawking,
The Four laws of black hole mechanics,
\hypersetup{urlcolor=blue}\href{https://doi.org/10.1007/BF01645742}{Commun. Math. Phys. \textbf{31}, 161-170 (1973)}.
\bibitem{Hawking:1975vcx}
S.~W.~Hawking,
Particle Creation by Black Holes,
\hypersetup{urlcolor=blue}\href{https://doi.org/10.1007/BF02345020}{Commun. Math. Phys. \textbf{43}, 199-220 (1975)}[erratum: Commun. Math. Phys. \textbf{46}, 206 (1976)].
\bibitem{Hawking:1982dh}
S.~W.~Hawking and D.~N.~Page,
Thermodynamics of Black Holes in anti-De Sitter Space,
\hypersetup{urlcolor=blue}\href{https://doi.org/10.1007/BF01208266}{Commun. Math. Phys. \textbf{87}, 577-588 (1983)}.
\bibitem{Witten:1998zw}
E.~Witten,
Anti-de Sitter space, thermal phase transition, and confinement in gauge theories,
\hypersetup{urlcolor=blue}\href{https://doi.org/10.4310/ATMP.1998.v2.n3.a3}
{Adv. Theor. Math. Phys. \textbf{2}, 505-532 (1998)},
[\hypersetup{urlcolor=magenta}\href{https://arxiv.org/abs/hep-th/9803131}
{arXiv:hep-th/9803131}].
\bibitem{Chamblin:1999tk}
A.~Chamblin, R.~Emparan, C.~V.~Johnson and R.~C.~Myers,
Charged AdS black holes and catastrophic holography,
\hypersetup{urlcolor=blue}\href{https://doi.org/10.1103/PhysRevD.60.064018}
{Phys. Rev. D \textbf{60}, 064018 (1999)},
[\hypersetup{urlcolor=magenta}\href{https://arxiv.org/abs/hep-th/9902170}
{arXiv:hep-th/9902170}].
\bibitem{Dolan:2010ha}
B.~P.~Dolan,
The cosmological constant and the black hole equation of state,
\hypersetup{urlcolor=blue}
\href{https://doi.org/10.1088/0264-9381/28/12/125020}
{Class. Quant. Grav. \textbf{28}, 125020 (2011)},
[\hypersetup{urlcolor=magenta}\href{https://arxiv.org/abs/1008.5023}
{arXiv:1008.5023}].
\bibitem{Kubiznak:2012wp}
D.~Kubiznak and R.~B.~Mann,
P-V criticality of charged AdS black holes,
\hypersetup{urlcolor=blue}\href{https://doi.org/10.1007/JHEP07(2012)033}{JHEP \textbf{07}, 033 (2012)},
[\hypersetup{urlcolor=magenta}\href{https://arxiv.org/abs/1205.0559}{arXiv:1205.0559}].
\bibitem{Gunasekaran:2012dq}
S.~Gunasekaran, R.~B.~Mann and D.~Kubiznak,
Extended phase space thermodynamics for charged and rotating black holes and Born-Infeld vacuum polarization,
\hypersetup{urlcolor=blue}\href{https://doi.org/10.1007/JHEP11(2012)110}{JHEP \textbf{11}, 110 (2012)},
[\hypersetup{urlcolor=magenta}\href{https://arxiv.org/abs/1208.6251}{arXiv:1208.6251}].
\bibitem{Dehyadegari:2016nkd}
A.~Dehyadegari, A.~Sheykhi and A.~Montakhab,
Critical behavior and microscopic structure of charged AdS black holes via an alternative phase space,
\hypersetup{urlcolor=blue}\href{https://doi.org/10.1016/j.physletb.2017.02.064}
{Phys. Lett. B \textbf{768}, 235-240 (2017)},
[\hypersetup{urlcolor=magenta}\href{https://arxiv.org/abs/1607.05333}
{arXiv:1607.05333}].
\bibitem{Dehyadegari:2017hvd}
A.~Dehyadegari and A.~Sheykhi,
Reentrant phase transition of Born-Infeld-AdS black holes,
\hypersetup{urlcolor=blue}\href{https://doi.org/10.1103/PhysRevD.98.024011}
{Phys. Rev. D \textbf{98}, no. 2, 024011 (2018)},
[\hypersetup{urlcolor=magenta}\href{https://arxiv.org/abs/1711.01151}
{arXiv:1711.01151}].
\bibitem{born1934quantum}
M. Born,
On the quantum theory of the electromagnetic field,
\hypersetup{urlcolor=blue}\href{https://doi.org/10.1098/rspa.1934.0010}
{Proc. Roy. Soc. Lond. A \textbf{143}, no. 849, 410-437 (1934)}.
\bibitem{born1934foundations}
M. Born and L. Infeld,
Foundations of the new field theory,
\hypersetup{urlcolor=blue}\href{https://doi.org/10.1098/rspa.1934.0059}
{Proc. Roy. Soc. Lond. A \textbf{144}, no. 852, 425-451 (1934)}.
\bibitem{ayon1999non}
E. Ayon-Beato and A. Garcia,
Non-singular charged black hole solution for non-linear source,
\hypersetup{urlcolor=blue}\href{https://doi.org/10.1023/A:1026640911319}
{Gen. Rel. Grav. \textbf{31}, no. 5, 629-633 (1999)},
[\hypersetup{urlcolor=magenta}\href{https://arxiv.org/abs/gr-qc/9911084}
{arXiv:gr-qc/9911084}].
\bibitem{ayon1999new}
E. Ayon-Beato and A. Garcia,
New regular black hole solution from nonlinear electrodynamics,
\hypersetup{urlcolor=blue}\href{https://doi.org/10.1016/S0370-2693(99)01038-2}
{Phys. Lett. B \textbf{464}, 25-29 (1999)},
[\hypersetup{urlcolor=magenta}\href{https://arxiv.org/abs/hep-th/9911174}
{arXiv:hep-th/9911174}].
\bibitem{de2002nonlinear}
V.~A.~De Lorenci, R.~Klippert, M.~Novello and J.~M.~Salim,
Nonlinear electrodynamics and FRW cosmology,
\hypersetup{urlcolor=blue}\href{https://doi.org/10.1103/PhysRevD.65.063501}
{Phys. Rev. D \textbf{65}, 063501 (2002)}.
\bibitem{irina2004regular}
I.~Dymnikova,
Regular electrically charged structures in nonlinear electrodynamics coupled to general relativity,
\hypersetup{urlcolor=blue}\href{https://doi.org/10.1088/0264-9381/21/18/009}
{Class. Quant. Grav. \textbf{21}, 4417-4429 (2004)},
[\hypersetup{urlcolor=magenta}\href{https://arxiv.org/abs/gr-qc/0407072}
{arXiv:gr-qc/0407072}].
\bibitem{corda2010removing}
C.~Corda and H.~J.~Mosquera Cuesta,
Removing black-hole singularities with nonlinear electrodynamics,
\hypersetup{urlcolor=blue}\href{https://doi.org/10.1142/S0217732310033633}
{Mod. Phys. Lett. A \textbf{25}, 2423-2429 (2010)},
[\hypersetup{urlcolor=magenta}\href{https://arxiv.org/abs/0905.3298}
{arXiv:0905.3298}].
\bibitem{corda2011inflation}
C.~Corda and H.~J.~Mosquera Cuesta,
Inflation from $R^2$ gravity: a new approach using nonlinear electrodynamics,
\hypersetup{urlcolor=blue}\href{https://doi.org/10.1016/j.astropartphys.2010.12.002}
{Astropart. Phys. \textbf{34}, 587-590 (2011)},
[\hypersetup{urlcolor=magenta}\href{https://arxiv.org/abs/1011.4801}
{arXiv:1011.4801}].
\bibitem{soleng1995charged}
H. H. Soleng,
Charged black points in general relativity coupled to the logarithmic U(1) gauge theory,
\hypersetup{urlcolor=blue}\href{https://doi.org/10.1103/PhysRevD.52.6178}
{Phys. Rev. D \textbf{52}, no. 10, 6178-6181 (1995)},
[\hypersetup{urlcolor=magenta}\href{https://arxiv.org/abs/hep-th/9509033}
{arXiv:hep-th/9509033}].
\bibitem{hassaine2007higher}
M. Hassaine and C. Martinez,
Higher-dimensional black holes with a conformally invariant Maxwell source,
\hypersetup{urlcolor=blue}\href{https://doi.org/10.1103/PhysRevD.75.027502}
{Phys. Rev. D \textbf{75}, no. 2, 027502 (2007)},
[\hypersetup{urlcolor=magenta}\href{https://arxiv.org/abs/hep-th/0701058}
{arXiv:hep-th/0701058}].
\bibitem{hendi2012asymptotic}
S. H. Hendi,
Asymptotic charged BTZ black hole solutions,
\hypersetup{urlcolor=blue}\href{https://doi.org/10.1007/JHEP03(2012)065}
{JHEP \textbf{1203}, 065 (2012)},
[\hypersetup{urlcolor=magenta}\href{https://arxiv.org/abs/1405.4941}
{arXiv:1405.4941}].
\bibitem{hendi2015thermodynamic}
S. H. Hendi and M. Momennia,
Thermodynamic instability of topological black holes with nonlinear source,
\hypersetup{urlcolor=blue}\href{https://doi.org/10.1140/epjc/s10052-015-3283-2}
{Eur. Phys. J. C \textbf{75}, no. 2, 54 (2015)},
[\hypersetup{urlcolor=magenta}\href{https://arxiv.org/abs/1501.04863}
{arXiv:1501.04863}].
\bibitem{Nwater}
C. S. Hudson,
Die gegenseitige Löslichkeit von Nikotin in Wasser,
\hypersetup{urlcolor=blue}\href{https://doi.org/10.1515/zpch-1904-4708}
{Zeitschrift für Physikalische Chemie \textbf{47}, no. 1, 113-115 (1904)}.
\bibitem{RPTreview}
T. Narayanan and A. Kumar,
Reentrant phase transitions in multicomponent liquid mixtures,
\hypersetup{urlcolor=blue}\href{https://doi.org/10.1016/0370-1573(94)90015-9}
{Physics Reports \textbf{249}, no. 3, 135-218 (1994)}.
\bibitem{KerrRPT}
N. Altamirano, D. Kubiznak and R. B. Mann,
Reentrant phase transitions in rotating anti-de Sitter black holes,
\hypersetup{urlcolor=blue}\href{https://doi.org/10.1103/PhysRevD.88.101502}
{Phys. Rev. D \textbf{88}, no.10, 101502 (2013)},
[\hypersetup{urlcolor=magenta}\href{https://arxiv.org/abs/1306.5756}
{arXiv:1306.5756}].
\bibitem{MRPT}
A. M. Frassino, D. Kubiznak, R. B. Mann and F. Simovic,
Multiple reentrant phase transitions and triple points in Lovelock
thermodynamics,
\hypersetup{urlcolor=blue}\href{https://doi.org/10.1007/JHEP09(2014)080}
{JHEP \textbf{1409}, 080 (2014)},
[\hypersetup{urlcolor=magenta}\href{https://arxiv.org/abs/1406.7015}
{arXiv:1406.7015}].
\bibitem{hairyRPT}
R. A. Hennigar and R. B. Mann,
Reentrant phase transitions and van der Waals behaviour for hairy black holes,
\hypersetup{urlcolor=blue}\href{https://doi.org/10.3390/e17127862}
{Entropy \textbf{17}, 8056 (2015)},
[\hypersetup{urlcolor=magenta}\href{https://arxiv.org/abs/1509.06798}
{arXiv:1509.06798}].
\bibitem{dSRPT}
D.~Kubiznak and F.~Simovic,
Thermodynamics of horizons: de Sitter black holes and reentrant phase transitions,
\hypersetup{urlcolor=blue}\href{https://doi.org/10.1088/0264-9381/33/24/245001}
{Class. Quant. Grav. \textbf{33}, no.24, 245001 (2016)},
[\hypersetup{urlcolor=magenta}\href{https://arxiv.org/abs/1507.08630}
{arXiv:1507.08630}].
\bibitem{Zou:2016sab}
D.~C.~Zou, R.~Yue and M.~Zhang,
Reentrant phase transitions of higher-dimensional AdS black holes in dRGT massive gravity,
\hypersetup{urlcolor=blue}\href{https://doi.org/10.1140/epjc/s10052-017-4822-9}
{Eur. Phys. J. C \textbf{77}, no.4, 256 (2017)},
[\hypersetup{urlcolor=magenta}\href{https://arxiv.org/abs/1612.08056}
{arXiv:1612.08056}].
\bibitem{MicroscopicRPT}
M.~Kord Zangeneh, A.~Dehyadegari, A.~Sheykhi and R.~B.~Mann,
Microscopic origin of black hole reentrant phase transitions,
\hypersetup{urlcolor=blue}\href{https://doi.org/10.1103/PhysRevD.97.084054}
{Phys. Rev. D \textbf{97}, no.8, 084054 (2018)},
[\hypersetup{urlcolor=magenta}\href{https://arxiv.org/abs/1709.04432}
{arXiv:1709.04432}].
\bibitem{Dehghani:2020blz}
A.~Dehghani, S.~H.~Hendi and R.~B.~Mann,
Range of novel black hole phase transitions via massive gravity: Triple points and $N$-fold reentrant phase transitions,
\hypersetup{urlcolor=blue}\href{https://doi.org/10.1103/PhysRevD.101.084026}
{Phys. Rev. D \textbf{101}, no.8, 084026 (2020)},
[\hypersetup{urlcolor=magenta}\href{https://arxiv.org/abs/2009.07980}
{arXiv:2009.07980}].
\bibitem{Liu:2022spy}
Y.~P.~Liu, H.~M.~Cao and W.~Xu,
Reentrant phase transition with a single critical point of the Hayward-AdS black hole,
\hypersetup{urlcolor=blue}\href{https://doi.org/10.1007/s10714-021-02886-0}
{Gen. Rel. Grav. \textbf{54}, no.1, 5 (2022)}.
\bibitem{Bai:2022vmx}
N.~C.~Bai, L.~Song and J.~Tao,
Reentrant phase transition in holographic thermodynamicsof Born-Infeld AdS black hole,
\hypersetup{urlcolor=blue}\href{https://doi.org/10.1140/epjc/s10052-024-12407-3}
{Eur. Phys. J. C \textbf{84}, no.1, 43 (2024)},
[\hypersetup{urlcolor=magenta}\href{https://arxiv.org/abs/2212.04341}
{arXiv:2212.04341}].
\bibitem{Frassino:2023wpc}
A.~M.~Frassino, J.~F.~Pedraza, A.~Svesko and M.~R.~Visser,
Reentrant phase transitions of quantum black holes,
\hypersetup{urlcolor=blue}\href{https://doi.org/10.1103/PhysRevD.109.124040}
{Phys. Rev. D \textbf{109}, no.12, 124040 (2024)},
[\hypersetup{urlcolor=magenta}\href{https://arxiv.org/abs/2310.12220}
{arXiv:2310.12220}].
\bibitem{vishveshwara1970scattering}
C. V. Vishveshwara,
Scattering of gravitational radiation by a Schwarzschild black-hole,
\hypersetup{urlcolor=blue}\href{https://doi.org/10.1038/227936a0}
{Nature \textbf{227}, 936–938 (1970)}.
\bibitem{press1971long}
W. H. Press,
Long wave trains of gravitational waves from a vibrating black hole,
\hypersetup{urlcolor=blue}\href{https://doi.org/10.1086/180849}
{Astrophys. J. Lett. \textbf{170}, L105-L108 (1971)}.
\bibitem{Chandrasekhar:1975zza}
S.~Chandrasekhar and S.~L.~Detweiler,
The quasi-normal modes of the Schwarzschild black hole,
\hypersetup{urlcolor=blue}\href{https://doi.org/10.1098/rspa.1975.0112}
{Proc. Roy. Soc. Lond. A \textbf{344}, 441-452 (1975)}.
\bibitem{Nollert:1999ji}
H. P. Nollert,
Quasinormal modes: the characteristic sound of black holes and neutron stars,
\hypersetup{urlcolor=blue}\href{https://doi.org/10.1088/0264-9381/16/12/201}
{Classical and Quantum Gravity \textbf{16}, 159-216 (1999)}.
\bibitem{Kokkotas:1999bd}
K. D. Kokkotas and B. G. Schmidt,
Quasinormal modes of stars and black holes,
\hypersetup{urlcolor=blue}\href{https://doi.org/10.12942/lrr-1999-2}
{Living Rev. Rel. \textbf{2}, 2 (1999)},
[\hypersetup{urlcolor=magenta}\href{https://arxiv.org/abs/gr-qc/9909058}
{arXiv:gr-qc/9909058}].
\bibitem{Konoplya:2002zu}
R.~A.~Konoplya,
On quasinormal modes of small Schwarzschild-anti-de Sitter black hole,
\hypersetup{urlcolor=blue}\href{https://doi.org/10.1103/PhysRevD.66.044009}
{Phys. Rev. D \textbf{66}, 044009 (2002)},
[\hypersetup{urlcolor=magenta}\href{https://arxiv.org/abs/hep-th/0205142}
{arXiv:hep-th/0205142}].
\bibitem{Konoplya:2002ky}
R.~A.~Konoplya,
Decay of charged scalar field around a black hole: Quasinormal modes of R-N, R-N-AdS black hole,
\hypersetup{urlcolor=blue}\href{https://doi.org/10.1103/PhysRevD.66.084007}
{Phys. Rev. D \textbf{66}, 084007 (2002)},
[\hypersetup{urlcolor=magenta}\href{https://arxiv.org/abs/gr-qc/0207028}
{arXiv:gr-qc/0207028}].
\bibitem{Cardoso:2003cj}
V.~Cardoso, R.~Konoplya and J.~P.~S.~Lemos,
Quasinormal frequencies of Schwarzschild black holes in anti-de Sitter space-times: A Complete study on the asymptotic behavior,
\hypersetup{urlcolor=blue}\href{https://doi.org/10.1103/PhysRevD.68.044024}
{Phys. Rev. D \textbf{68}, 044024 (2003)},
[\hypersetup{urlcolor=magenta}\href{https://arxiv.org/abs/gr-qc/0305037}
{arXiv:gr-qc/0305037}].
\bibitem{Wang:2005vs}
B. Wang,
Perturbations around black holes,
\hypersetup{urlcolor=blue}\href{https://doi.org/10.1590/S0103-97332005000700002}
{Braz. J. Phys. \textbf{35}, 1029-1037 (2005)},
[\hypersetup{urlcolor=magenta}\href{https://arxiv.org/abs/gr-qc/0511133}
{arXiv:gr-qc/0511133}].
\bibitem{Fernando:2005bc}
S. Fernando and C. Holbrook,
Stability and quasi normal modes of charged black holes in Born-Infeld gravity,
\hypersetup{urlcolor=blue}\href{https://doi.org/10.1007/s10773-005-9024-9}
{Int. J. Theor. Phys. \textbf{45}, 1630-1645 (2006)},
[\hypersetup{urlcolor=magenta}\href{https://arxiv.org/abs/hep-th/0501138}
{arXiv:hep-th/0501138}].
\bibitem{Konoplya:2011qq}
R. A. Konoplya and A. Zhidenko,
Quasinormal modes of black holes: from astrophysics to string theory,
\hypersetup{urlcolor=blue}\href{https://doi.org/10.1103/RevModPhys.83.793}
{Rev. Mod. Phys. \textbf{83}, 793-836 (2011)},
[\hypersetup{urlcolor=magenta}\href{https://arxiv.org/abs/1102.4014}
{arXiv:1102.4014}].
\bibitem{Mahapatra:2016dae}
S.~Mahapatra,
Thermodynamics, Phase Transition and Quasinormal modes with Weyl corrections,
\hypersetup{urlcolor=blue}\href{https://doi.org/10.1007/JHEP04(2016)142}
{JHEP \textbf{04}, 142 (2016)},
[\hypersetup{urlcolor=magenta}\href{https://arxiv.org/abs/1602.03007}
{arXiv:1602.03007}].
\bibitem{Lin:2024ubg}
J.~Lin, M.~Bravo-Gaete and X.~Zhang,
Quasinormal modes, greybody factors, and thermodynamics of four dimensional AdS black holes in critical gravity,
\hypersetup{urlcolor=blue}\href{https://doi.org/10.1103/PhysRevD.109.104039}
{Phys. Rev. D \textbf{109}, no. 10, 104039 (2024)},
[\hypersetup{urlcolor=magenta}\href{https://arxiv.org/abs/2401.02045}
{arXiv:2401.02045}].
\bibitem{Ribeiro:2025amp}
M.~R.~Ribeiro, E.~C.~Ribeiro, K.~Lin and E.~Abdalla,
Universality in quasinormal modes of a magnetized black hole,
\hypersetup{urlcolor=blue}\href{https://doi.org/10.1103/2bc9-zb5b}
{Phys. Rev. D \textbf{113}, no. 12, 124029 (2026)},
[\hypersetup{urlcolor=magenta}\href{https://arxiv.org/abs/2512.08116}
{arXiv:2512.08116}].
\bibitem{Liu:2014gvf}
Y. Liu, D. C. Zou and B. Wang,
Signature of the Van der Waals like small-large charged AdS black hole phase transition in quasinormal modes,
\hypersetup{urlcolor=blue}\href{https://doi.org/10.1007/JHEP09(2014)179}
{JHEP \textbf{09}, 179 (2014)},
[\hypersetup{urlcolor=magenta}\href{https://arxiv.org/abs/1405.2644}
{arXiv:1405.2644}].
\bibitem{Priyadarshinee:2023exb}
S. Priyadarshinee,
Quasinormal mode of dyonic hairy black hole and its interplay with phase transitions,
\hypersetup{urlcolor=blue}\href{https://doi.org/10.1140/epjp/s13360-024-05044-y}
{Eur. Phys. J. Plus \textbf{139}, no. 3, 258 (2024)},
[\hypersetup{urlcolor=magenta}\href{https://arxiv.org/abs/2308.05719}
{arXiv:2308.05719}].
\bibitem{Guo:2024jhg}
Y. Guo, H. Xie, Y. G. Miao,
Signal of phase transition hidden in quasinormal modes of regular AdS black holes,
\hypersetup{urlcolor=blue}\href{https://doi.org/10.1016/j.physletb.2024.138801}
{Phys. Lett. B \textbf{855}, 138801 (2024)},
[\hypersetup{urlcolor=magenta}\href{https://arxiv.org/abs/2402.10406}
{arXiv:2402.10406}].
\bibitem{Hou:2025bli}
Z.~Y.~Hou, Y.~Q.~Lei and X.~H.~Ge,
Quasinormal modes and nonlinear electrodynamics in black hole phase transitions,
\hypersetup{urlcolor=blue}\href{https://doi.org/10.1103/f333-dnc6}
{Phys. Rev. D \textbf{113}, no. 2, 024060 (2026)},
[\hypersetup{urlcolor=magenta}\href{https://arxiv.org/abs/2508.00404}
{arXiv:2508.00404}].
\bibitem{Rosh:2025}
F. Roshan Bakhsh, M. Kord Zangeneh and D. Afshar,
The isocharge phase transition signature in quasi-normal modes of charged anti-de Sitter black holes,
\hypersetup{urlcolor=blue}\href{https://doi.org/10.22055/jrmbs.2024.19702}
{Journal of Research on Many-body Systems \textbf{14}, no. 3, 54-65 (2025)}.
\bibitem{Berti:2009kk}
E.~Berti, V.~Cardoso and A.~O.~Starinets,
Quasinormal modes of black holes and black branes,
\hypersetup{urlcolor=blue}\href{https://doi.org/10.1088/0264-9381/26/16/163001}
{Class. Quant. Grav. \textbf{26}, 163001 (2009)},
[\hypersetup{urlcolor=magenta}\href{https://arxiv.org/abs/0905.2975}
{arXiv:0905.2975}].
\bibitem{Jaramillo:2021tmt}
J.~L.~Jaramillo, R.~Panosso Macedo and L.~A.~Sheikh,
Gravitational wave signatures of black hole quasinormal mode instability,
\hypersetup{urlcolor=blue}\href{https://doi.org/10.1103/PhysRevLett.128.211102}
{Phys. Rev. Lett. \textbf{128}, no. 21, 211102 (2022)},
[\hypersetup{urlcolor=magenta}\href{https://arxiv.org/abs/2105.03451}
{arXiv:2105.03451}].
\end{thebibliography}
\end{document}